\documentclass{kluwer}    

\newdisplay{guess}{Conjecture}
\usepackage{graphicx}
\begin{document}                                                                                   
\begin{article}
\begin{opening}         
\title{An update on Archeops: flights and data products.} 
\author{Jacques Delabrouille$^1$, Philippe Filliatre$^1$,\\
and the Archeops Collaboration}  
\runningauthor{The Archeops collaboration}
\runningtitle{The Archeops experiment}
\institute{$^1$ PCC --- Coll\`ege de France, 11 place Marcelin Berthelot, F--75231 Paris cedex 05 }
\date{April 10, 2003}

\begin{abstract}
     Archeops is a balloon--borne instrument dedicated to measuring
cosmic microwave background (CMB) temperature anisotropies at high
angular resolution ($\sim 8$ arcminutes) over a large fraction ($\sim
30$\%) of the sky in the millimetre domain.  The general design is
based on Planck High
Frequency Instrument (HFI) technology. Bolometers cooled to 0.1~K scan
the sky in total power mode along large circles at constant elevation.
Archeops is designed to observe a complete annulus on the sky covering all right ascensions between about 25 and 55 degrees during the course of a 24--hour Arctic--night balloon flight,  in four frequency bands
centered at 143, 217, 353 and 545~GHz. We describe the Archeops flights and the data products obtained during the three successful flights from Trapani (Sicily) to
Spain in July 1999, and from Kiruna (Sweden) to Russia in January 2001 and February 2002. We discuss present Archeops results and the future use of Archeops data. 

\end{abstract}
\end{opening}           

\section{Introduction}  

The anisotropies of the Cosmic Microwave Background (CMB) are a
goldmine of cosmological information. They tell us about the state of
the Universe at the age of a few hundred thousand years
(redshift $z \sim 1000$), when the primeval plasma combined into neutral atoms
and became transparent, unveiling the small fluctuations that would eventually form the
large--scale structure of the present--day Universe.  These small
inhomogeneities provide the cosmologist with a (relatively) direct link
between observable and theoretical quantities, so that accurate measurements 
of the anisotropies can be used to constrain fundamental parameters of the standard
Big Bang cosmology, at least within the adopted theoretical framework.  
For the case of inflation--generated
perturbations, this approach directly constrains the spatial
geometry of the Universe, the physical particle densities (such as the
baryon or dark matter density), the form of the primeval
perturbation spectrum (scalar and tensor modes), and the
ionization history since recombination.  A large international effort
has set as its goal high precision measurements of the CMB
anisotropies, employing numerous ground--based, balloon and
space--borne projects; among the latter WMAP, launched by NASA in
2001, and Planck to be launched in 2007 by ESA.

The Archeops experiment is, with the WMAP space--borne mission, one of the few sensitive instruments mapping a large fraction of the sky in a large range of observing wavelengths. In addition to measuring the CMB power spectrum, Archeops data provides a survey of astrophysical emissions between 150 and 550 GHz over more than 30\% of the sky. The corresponding multifrequency maps can be used for a variety of applications.

\section{The Archeops experiment}

\subsection{Mission objectives}

At the time of Archeops preliminary design studies in 1997-1998, several sensitive ground based and balloon--borne instruments dedicated to middle--range CMB power spectrum measurement were already in operating status. Most of these experiments observed small regions of the sky as free as possible from galactic foregrounds, yielding measurements of the spectrum on degree angular sizes and smaller, at $\ell > 100$. On the other hand, the COBE-DMR space--borne instrument had yielded a measurement of the CMB power spectrum at very large scales, on angular scales larger than 10 degrees or so. A sensitive instrument capable of measuring the spectrum on a wide angular scale range from $\ell \simeq 20$ to the second acoustic peak was missing.

A few experiments were designed for this purpose, among which Archeops, which measured the first acoustic peak from $\ell = 20$ to $\ell = 500$ quite accurately in 2002 and the WMAP mission, which provided us early 2003 with a very accurate measurement of the CMB power spectrum, from the dipole to the rise of the third acoustic peak.

WMAP and Archeops together provide us with sensitive maps of a large common fraction of sky of more than 30\% in nine frequency channels ranging from 22 to 550 GHz.

\subsection{Concept}

Archeops is designed as an intermediate angular scales CMB mission based on the technology developed for the Planck HFI. Spider--Web bolometers, cooled down to 100 mK thanks to an open cycle $^3$He--$^4$He dilution fridge, are placed at the focal plane of a 1.3--meter gregorian off-axis telescope pointed at a 41--degree elevation. For sensitivity in a low background environment as well as minimization of atmospheric contamination, the instrument is flown on a stratospheric balloon. The beams of the instrument are scanned on the sky by rotating the gondola around a vertical axis during the flight. 

The mission objectives of the Archeops instrument require a large sky coverage for low $\ell$ sensitivity, and a 10--arcminute angular resolution for sensitivity to small angular scales. In addition, the necessity to cover a large fraction of the sky requires a long duration night-time flight, avoiding systematics and stray signals from the Sun and the Moon while spending enough time in each sub--degree pixel. For science flights, Archeops is launched by CNES from the Esrange base near Kiruna in Sweden during the polar arctic night. Typical wind speeds, power supply from batteries, and the autonomy of the Archeops cryostat permit flight durations of up to 36 hours.

\subsection{Observing strategy}

The Archeops scanning strategy consists in spinning the gondola at 2 rpm while letting the sky drift overheads. Circular scan trajectories shift on the sky during the flight, yielding a large sky coverage of interconnected scans.

Theoretical studies carried on the analysis of CMB data on circular scans \cite{Delabrouille:1998a,Delabrouille:1998b} demonstrate that it is possible to remove most of the unwanted effects of timeline slow drifts for such a scan strategy. When scans are well connected, the redundancy provided by scan crossings permit the removal of slow drifts in a map--making step. When they are not, a low-cut filter is applied to the data and accounted for in the spectral estimation step.

\section{Instrument}

\subsection{Detectors}

The Archeops detectors are spider--web bolometers similar in design to those of Planck HFI, Boomerang and MAXIMA. These detectors have been developed by JPL/Caltech as part of the Planck HFI development.  These bolometers have high
responsivity, low NEP, fast speed of response and a low cross section
to cosmic ray hits. The bolometers are made by micro-machining silicon
nitride to leave a self-supported, metallized spider web mesh absorber.
A neutron--transmutation--doped Ge:Ga thermistor is 
bonded at the web center to sense the temperature rise which results
from the absorption of radiation \cite{Bock:1996}, \cite{Mauskopf:1997}.

\subsection{Optics}
The Archeops telescope is a two mirror, off-axis, tilted Gregorian
telescope consisting of a parabolic primary and an elliptical
secondary. Radiation from the
telescope is focussed into the entrance of a back--to--back horn pair.
A lens at the exit aperture of the
second horn creates a beam--waist where wavelength selective filters
are placed.  A second lens on the front of the third horn
maintains beam control and focuses the radiation onto the spider
bolometer placed at the exit aperture.

\subsection{Cooling system}
The focal plane is cooled to 100~mK by an open cycle dilution
refrigerator designed at the CRTBT (Centre de Recherche sur les Tr\`es
Basses Temp\'eratures) in Grenoble (France) \cite{benoit:1994a,benoit:1994b}.  The dilution stage is placed in a low
temperature box placed on a liquid Helium reservoir at 4.2~K.
A heat exchanger using exhaust
vapours from the helium tank maintains the horns near 10K.

\subsection{Star sensor and pointing reconstruction}
A custom star sensor, fast enough to work on a payload rotating at 2--3 rpm, has been developed for pointing reconstruction. 
A linear array of 46 sensitive photodiodes,
placed in the focal plane of a 40~cm diameter, 1.8~m focal length
parabolic optical mirror, scan the sky along a circle
at an elevation of $41^\circ \pm 1^\circ$. An algorithm treats independent photodiode data
streams to detect bright star triggers, and matches the detected stars against a star catalogue for pointing reconstruction.

In addition to the fast star sensor, gyroscopes give information about balloon pendulation, and a magnetometer gives phase information from the flux of the earth magnetic field. Both informations can be used to help the pointing reconstruction algorithm.

The constant offset between bolometer and stellar sensor pointing directions on the sky is calibrated on Jupiter crossings.

\section{Archeops Flights}

The Archeops experiment has been launched successfully three times. A first test flight was made from Trapani in july 1999 with an early version of the experiment \cite{papier_trapani}. Two successful scientific flights were made from Kiruna in January 2001 and February 2002. A summary of in-flight performance and data set characteristics is given on table \ref{tab:data-sets}.

\begin{table*}
\caption[]{Summarized in-flight performance and data set characteristics. Sensitivities are average values estimated after removing low frequency drifts, and assuming an even distribution of the integration time over the whole observed region.}
\label{tab:data-sets}
\begin{tabular}{lccccc}
\hline
flight & frequency & \# channels & FWHM & $\mu$K.deg$^{-1}$ & $f_{\rm sky}$ \\
\hline
Trapani T99 & 143 GHz & 2 & 11' & 110 & 19\% \\
Trapani T99 & 217 GHz & 1 & 12' & 96 & 19\% \\
Trapani T99 & 353 GHz & 1 & 10' & 374 & 19\% \\
\hline
Kiruna KS1 & 143 GHz & 6 & 10' & 57 & 23\% \\
Kiruna KS1 & 217 GHz & 5 & 10' & 51 & 23\% \\
Kiruna KS1 & 353 GHz & 5 & 10' & 384 & 23\% \\
Kiruna KS1 & 545 GHz & 2 & 10' & 8337 & 23\% \\
\hline
Kiruna KS3 & 143 GHz & 6 & 10' & 33 & 30\% \\
Kiruna KS3 & 217 GHz & 7 & 10' & 48 & 30\% \\
Kiruna KS3 & 353 GHz & 6 & 10' & 437 & 30\% \\
Kiruna KS3 & 545 GHz & 1 & 10' & 3417 & 30\% \\
\hline
\end{tabular}
\end{table*}

\subsection{The test flight}
The test flight used a reduced focal plane (six detectors only, four of which worked satisfactorily). The main objective was to test the general behaviour of the instrument, and in particular the dilution fridge which had never been flown before. This test was critical not only for future Archeops scientific flights, but also as a feasibility test for the Planck HFI.

Launch of the balloon was provided by ASI (Italian Space Agency) base
near Trapani (Sicily) in the evening of the 17th July 1999. The flight started at 19.37 UT with the launch from the Trapani launch pad, and lasted until separation of the gondola at 12.32 UT the next day. The scientifically useful data ranges from 23.45 UT on the 17th, when the instrument operation parameters were fully adjusted  at the nominal flight altitude around 41 km, until sunrise at 4.00 UT the next day (about four hours of useful data).

\subsection{Scientific flights}

\subsubsection{KS1 flight}

The first scientific flight of Archeops took place on January 29th 2001 from the Swedish balloon base of Esrange near Kiruna, at a latitude of about 68$^\circ$ north, where the north polar night provides, in principle, long night time observation conditions. 

Unfortunately, the polar vortex was not as favorable as expected from the average meteorological conditions of the previous years. The balloon could not be launched succesfully before end of january, quite late in the season. In addition, strong wind conditions limited the flight to the relatively low float altitude of 32 km, and to about 7.5 hours of useful data. The nominal altitude (of about 35 km) could not be reached because of the unfavorable northbound direction of higher altitude winds, which resulted in a higher thermal load and strong spurious emissions of atmospheric origin. Nonetheless, the collected data is of high scientific interest, as discussed later.

During the flight, the focal plane was kept to a temperature of about 90 mK. The instrument performed satisfactorily for the whole flight duration.

\subsubsection{KS3 flight}

The experiment was launched again from Esrange on the 7th of February. The flight started at 13h40 UT and lasted until 10h40 UT the next day, when the payload was separated from the balloon short after crossing the Ienissei river near Noril'sk in Siberia. Excellent night-time data was collected between 15.3 UT the 7th and 3.0 UT the 8th. During the corresponding 11.7 hours, the payload flew at an average altitude of 35 km in nominal operating conditions.

\section{The data}

\subsection{Data description}

\subsubsection{Data products}
Several data products are being prepared from Archeops data : timelines, maps of the sky, maps of components, spatial power spectra.

Scientific data is collected in the form of timelines (Time-ordered information or TOI) for all bolometers, in a total power mode. For the last flight, which produced the best data-set for CMB studies, timelines are made of about $7\times 10^{6}$ samples per detector, collected  at the acquisition frequency of 152 Hz, which corresponds to a sampling period of about 3.6 arcminutes along the scans on the sky. Cleaned timelines, associated to pointing information, are produced subsequently, as described in section \ref{sec:processing}. 

\begin{figure}[t]
\begin{center}
\includegraphics[height=\textwidth,angle=90]{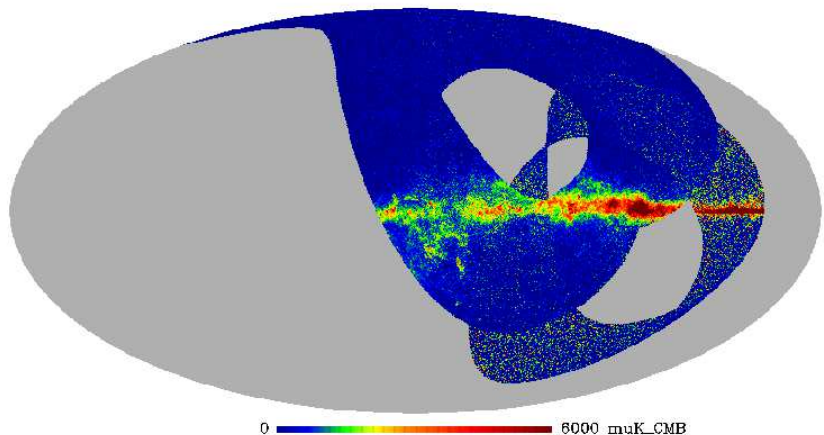}
\caption{An illustrative map obtained from combining the 217 GHz Archeops data from the Trapani flight (the noisiest part) and the two Kiruna flights. The coverage of the second Kiruna flight encompasses entirely the coverage of the first one. A robust combination of all Archeops data sets remains to be done.}
\label{maps}
\end{center}
\end{figure}

The data from each detector can be turned into a part-sky HEALPix map for further processing. Several methods for doing so are being developed. Combined maps are obtained in each frequency channels by weighted averaging of single detector maps. An illustration of the Archeops sky coverage is shown in figure \ref{maps}.

Finally, component maps are estimated by combining the data from all good channels, and the CMB power spectrum is estimated by several independent methods using the various available maps.

\subsection{Sensitivity and noise spectra}

The sensitivity of Archeops is estimated on the high-frequency part of the timelines, where the power is dominated by white noise contributions. Because of slow drifts, the effective sensitivity on maps can be somewhat worse when the map--making method does not efficiently subtract long--term drifts. Such long term drifts contribute essentially at frequencies below 1 Hz, which represents less than 2\% of the useful bandwidth.

\begin{figure}[t]
\begin{center}
\includegraphics[height=\textwidth,angle=90]{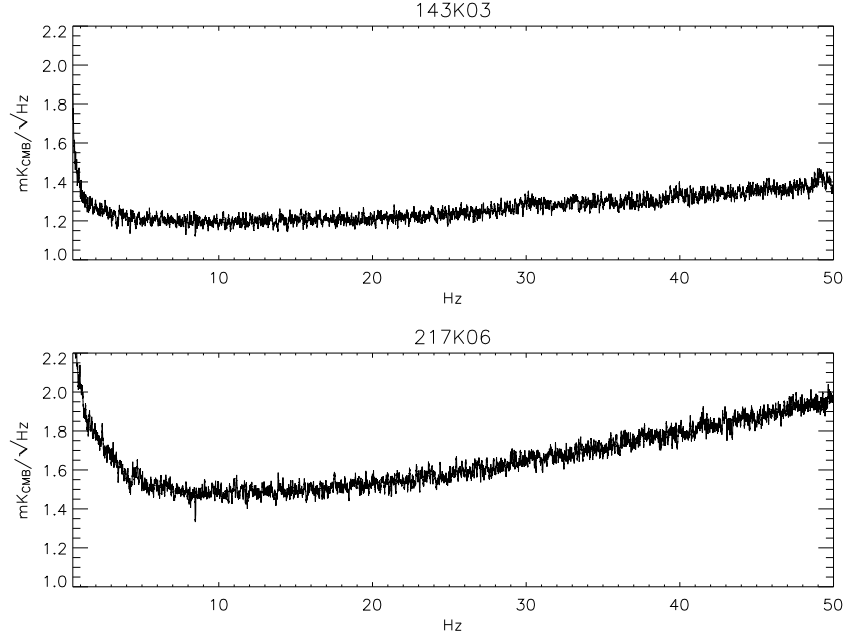}
\caption{The noise power spectra of the two bolometers used in the analysis, 143K03 and 217K06. The timeline has been deconvolved from the effect of the bolometer time constant for a flat response over the frequencies.}
\label{fig:noise_spectra}
\end{center}
\end{figure}

The estimated noise power spectra of two of the best bolometers are shown in figure \ref{fig:noise_spectra}. 

\subsection{Systematics}

Various systematic effects exist in the Archeops data. They are listed below, starting from those which affect essentially very low frequencies to those affecting high frequencies in the bolometer timelines.

\begin{enumerate} 
\item A slow cooling of the working temperature of the bolometers while the instrument observes the cold sky at cruise altitude induces a slow increase in the bolometer sensitivity as time passes.
\item Balloon pendulations and altitude fluctuations generate a fluctuation in airmass seen by the bolometers, which induces low-frequency parasitic signals.
\item Temperature fluctuations of the various cold stages induce fluctuations in the amount of internal stray light impinging the detectors.
\item Perturbations and inhomogeneities in the atmosphere composition and temperature (e.g. ozone clouds), varying with time and with pointing direction, induce unwanted radiation pick-up at intermediate frequencies between a fraction of and a few times the spinning frequency of the gondola.
\item Cosmic ray hits on the detectors induce glitches in the data.
\item Occasional noise bursts of unknown origin are observed on some detector channels.
\item Bursts of microphonic noise or electromagnetic interference are seen on some detectors and on temperature--monitoring TOI from thermometers attached to the 100 $\mu K$ stage of the cryostat.
\end{enumerate}

Most of these unwanted signals can be identified and subtracted from the data, either by decorrelation methods, or by filtering. The remaining of the bad data is simply discarded (e.g. data flagged as affected by a cosmic ray impact).

\subsection{Data processing}\label{sec:processing}

\subsubsection{Methodology}

The present pipeline for processing Archeops timelines can be summarized as follows:

\begin{enumerate}
\item Low frequency noise of electrical origin in the timelines is avoided by modulating the signal with a square wave modulation in the data acquisition step. The first step of data processing is the synchronous demodulation of the data.
\item The small change of the calibration coefficients due to the long term drift of the temperature of the bolometer plate during the flight is corrected for using a model of the response of the bolometer as a function of working temperature, computed from bias--current curves obtained during ground calibration.
\item Corrupted data of various types (holes, glitches, noise bursts, unsecure pointing, ...), are identified and flagged with an automatic algorithm followed by visual inspection for confirmation. Such bad data represents about 3\% of the total data.
\item Very slow drifts correlated to drifts of the temperature of the various stages of the crysotat and to airmass are removed by decorrelation.
\item Photometric calibration is performed on the CMB dipole.
\item High frequency systematics correlated between scientific and housekeeping data, due probably to microphonics and/or electromagnetic interference, are removed from the scientific data by decorrelation in a time-frequency analysis.
\item The geometry of the focal plane and bolometer time constants are measured on the signal observed from bright planets (Jupiter, Saturn).
\item Bolometer timelines are deconvolved from bolometer time constants.
\item Independent photometric calibrations are performed on Jupiter and on the galactic plane, and compared to calibrations obtained on the CMB dipole.
\item Noise power spectra for each bolometer are estimated.
\item The quality of the data of each bolometer in the 143 and 217 GHz frequency channels is evaluated according to a number of criteria (noise level, noise stationnarity).
\end{enumerate}

\subsubsection{Cleaning}

\begin{figure}[t]
\begin{center}
\includegraphics[height=\textwidth,angle=90]{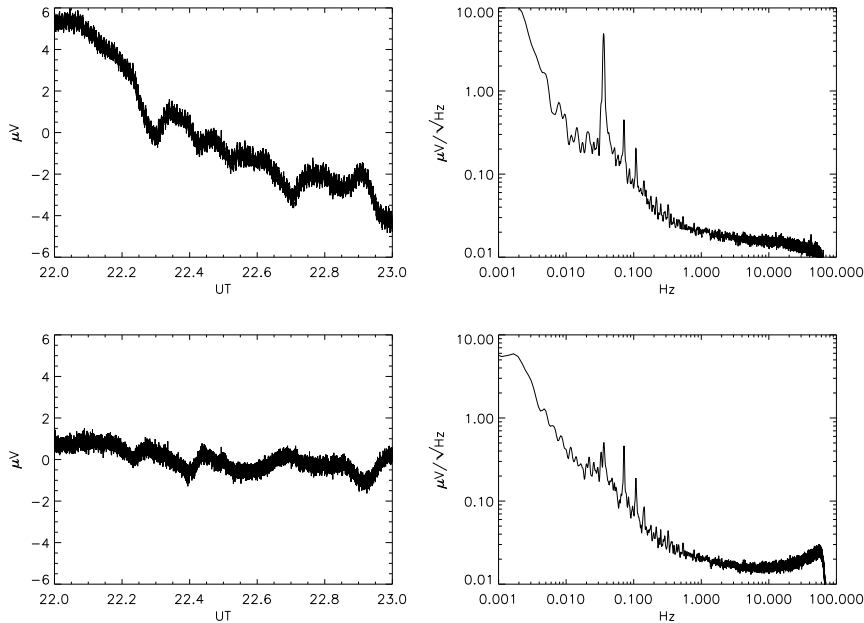}
\caption{Left column : the TOI of the best 143 GHz bolometer in time domain (left), and the Amplitude Spectral Density in the Fourier domain (right). Top : before cleaning; bottom: after processing for low and high frequency effects, time constant deconvolution, and dipole calibration.}
\label{fig:cleaning}
\end{center}
\end{figure}

The cleaning pipeline, described in more detail elsewhere \cite{papier_processing}, was made in successive development steps which refine both the measurement model and the processing algorithms. The timeline and power spectrum of a detector signal before and after cleaning are shown on figure \ref{fig:cleaning}.

\subsubsection{Map making}

None of the Archeops flights was long enough to yield a well connected sky coverage with high redundancy. For each flight, scan crossings are located in limited regions of the sky. Slow drifts of various origins are present in the data. Hence, standard map--making methods are not so efficient at removing all the slow drifts. Five independent methods have been and are being developed for Archeops data processing. Although none of the methods used is fully satisfactory yet for producing releasable maps, they permit the further analysis of the maps for the estimation of the spatial power spectrum of the CMB with a Monte--Carlo correction of the biases introduced (see \ref{sec:C-ell}).

The first method consists in producing sub-optimal maps by simple noise-weighted averaging into pixel bins of Archeops time samples using TOIs in which the lowest frequency modes have been filtered out by a high--pass filter. This first method has several drawbacks. First, it removes some power from the sky emission itself. Second, it does not remove the same sky power for all detectors, as all detectors do not point simultaneously at the same point on the sky (the pointing of a detector depends on its location in the focal plane). In addition, the filtering effect impacts CMB maps anisotropically, differently along scan direction and along cross-scan direction. This has impact on further analysis as for instance computing correlations between Archeops and other maps, or on power spectrum estimates, or on component separation from multifrequency observation.

The second method is the MAPCUMBA multiscale iterative map-making algorithm of Dor\'{e} et al. \shortcite{dore}. This method is not fully efficient here because of a lack of redundancy in the Archeops data and of scan-synchronous systematics. It does not suppress quite satisfactorily noise and systematics at intermediate scales.

The third method is the MIRAGE map-making algorithm of Yvon et al. \shortcite{mirage}, which combines both an efficient filtering and an iterative map-making scheme. This map--making has the drawback of removing some of the small--scale power in the maps in a way which depends on the number of iterations in the map--making part.

The fourth method uses the prior that the map spatial power spectrum (or two--dimensional correlation function) in high galactic latitude local regions should be isotropic. The method minimizes a weighted sum of local cross-scan variance to identify and remove additional cross-scan contribution due to residual striping. This method cleans the maps reasonably well on small scales, but disturbs somewhat the larger scale fluctuations.

Finally, the fifth method consists in using WMAP CMB templates \cite{wmap} and dust templates from Schlegel et al. \shortcite{schlegel} as additional prior information to constrain the shape of slow drifts in Archeops data and for subtracting them from the Archeops TOIs before reprojection. The method seems to perform quite well, but resulting maps depend to some extent (to be quantified) on the underlying model of the emission laws of CMB and dust across frequencies (including uncertainties in the Archeops calibration).

Map--making with Archeops is still the object of investigation.

\subsubsection{$C_\ell$ estimation}

\label{sec:C-ell}

The present Archeops measurement of the spatial power spectrum of CMB anisotropies \cite{papier_cl} is obtained using the first map--making method and the MASTER approach of Hivon et al \shortcite{master} to account for the biases that this choice introduces in the $C_{\ell}$ estimation. We used the timelines of the two best Archeops detectors, with additionnal processing :
\begin{itemize}
\item the timelines are decorrelated from a smoothed version of the 545 GHz bolometer timeline and a synthetic dust timeline obtained from the Schlegel et al. dust model \shortcite{schlegel}, to remove the contribution of the galactic dust and atmospheric signal ;
\item a band-pass filter is applied between 0.3 and 45 Hz, corresponding to about 30 degrees and 15 arcminutes scales, respectively~; the high-pass filter removes remaining atmospheric and galactic contamination, the low-pass filter suppresses non-stationary high-frequency noise.
\item only the data on the north side of $b=30^{\rm o}$, where the galactic contamination is low, are considered.
\end{itemize}
The timelines are then coadded into a single map, with a $1/\sigma^2$ weighting per 15 arcminute pixel (Healpix parameter nside=256, the total number of pixels being around $10^5$). The power spectrum of this map, the so-called pseudo-power spectrum, is corrected by the Master method to obtain the maximum-likelihood estimator of the true power spectrum~:
\begin{itemize}
\item the incompleteness of the sky coverage is corrected by the application of the coupling mode matrix, computed analytically, and properly rebinned to allow its inversion ;
\item the finite beam resolution and pixel size are corrected by the analytically computed window function ;
\item for the whole timeline processing, including the pipeline and the filtering done for the map-making, a transfer function in multipole space is computed by a Monte-Carlo method using simulated CMB timelines ; it has been checked by Hivon et al. \shortcite{master}, and by the Archeops collaboration, that the transfer function depends only little to the chosen set of cosmological parameters ;
\item the $C_{\ell}$ power spectrum is de-biased from the noise contribution by subtracting a noise angular power spectrum estimate obtained by a Monte Carlo approach using a robust estimation of the noise spectrum on the timelines, accounting for projection/deprojection effects \cite{amblard}. Noise spectra for the best Archeops data channel in the KS3 flight are displayed in figure \ref{fig:noise_spectra}.
\item error bars, similarly, are estimated by Monte Carlo simulations.
\end{itemize}

\begin{figure}[t]
\begin{center}
\includegraphics[scale=0.5,angle=90]{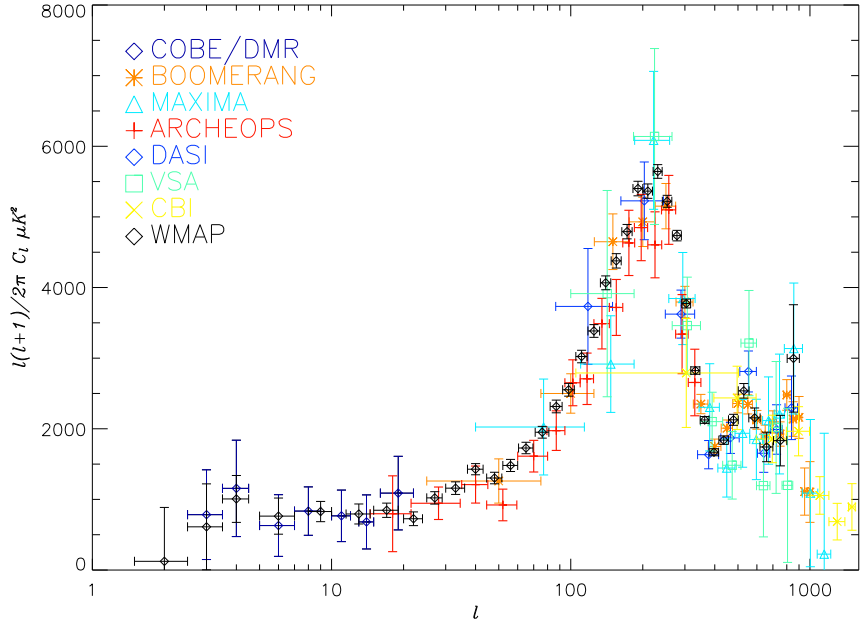}
\caption{The Archeops power spectrum compared with results of COBE, Boomerang, Maxima, Dasi, VSA, CBI and WMAP.}
\label{fig.cl}
\end{center}
\end{figure}

The CMB poser spectrum is estimated in 16 nearly-independant bands, shown on figure \ref{fig.cl}, alongside to other experiment results, COBE \cite{cobe}, Boomerang \cite{boomerang}, Maxima \cite{maxima}, Dasi \cite{dasi}, VSA \cite{vsa}, CBI \cite{cbi} and WMAP \cite{wmap}. Archeops was the first experiment to connect the COBE measurements of the Sachs-Wolfe plateau to the first acoustic peak, with small error bars and good sampling in the rise of the peak. 

This power spectrum is supported by a series of robustness tests. Changes of the cut frequencies of the pass band filter used for the map, of galactic cut, of binning in $\ell$ space do not modify the spectrum. Jack-knife tests shows agreement between the first and second halves of the flight, left and right part of the map. Two difference maps can be constructed : the difference of the two bolometers, and a map constructed by inverting the sign of the timeline from one Archeops rotation to another, so that the map must contain only noise. In both cases, the recovered power spectra are consistent with zero. Systematic effects, including a bad calculation of the bolometer time constants and contribution of atmosphere and Galactic dust, have been estimated to be small in respect to the statistical error bars. Finally, the power spectrum has been computed by two independant approaches, the SpICE \cite{spice} method, and a spectral matching method \cite{patanchon}. Both additional methods give compatible results, and will be pushed further for future improvements on the Archeops $C_{\ell}$.

\subsubsection{Component separation}

Archeops data is available at four observing frequencies: 150, 217, 353 and 545 GHz.
Such multi-frequency observations permit to identify and separate astrophysical 
emissions contributing to the total emission.
A component separation based on spectral matching \cite{delabrouille02} is being implemented on Archeops maps. This method permits to identify the CMB component as a component seen consistently across Archeops channels with the proper color, and yields both a measurement of the CMB power spectrum and a Wiener--filtered version of the Archeops CMB map.

\section{Science with Archeops}

\subsection{Parameter estimation}

\subsubsection{Method}

The useful data set for cosmological parameter estimation includes CMB experiments (COBE, Boomerang, Dasi, Maxima, VSA, CBI and Archeops, a combination thereafter referred as Archeops+CMB), and HST results on the Hubble constant \cite{hubble}. We constructed a database of $4.5\times 10^{8}\,C_{\ell}$ considering only inflationary models with adiabatic fluctuations. The free parameters are $\Omega_{tot}$, $\Omega_{\Lambda}$, $\Omega_{b}h^{2}$, $h$, $n_{s}$, $Q$ and $\tau$, whereas $r$ is set to zero and the dark matter is considered to be only cold.\\
Cosmological parameter estimation relies upon the knowledge of the likelihood function {\cal L} of each band power estimate. Current Monte-Carlo methods for the extraction of $C_{\ell}$ provide the distribution function {\cal D} of these power estimates. The analytical approach described in \cite{douspis} and \cite{bartlett} allows to construct {\cal L} from {\cal D}, and has been proven to be equivalent to performing a full likelihood analysis on the maps. Using {\cal L}, we calculate the likelihood of any of the cosmological model in the database and maximize it over the 7\% calibration uncertainty of each experiment.

\subsubsection{Results}

\begin{table}
\caption{Cosmological parameter constraints from 1--Archeops+CMB and 2--Archeops+CMB+HST datasets. Upper and lower limite are given for 68\% confidence level. The central values are given by the mean of the likelihood.}
\label{table_param}
\begin{tabular}{|c|cccccc|}
\hline
Data set&$\Omega_{tot}$&$n_{s}$&$\Omega_{b}h^{2}$&$h$&$\Omega_{\Lambda}$&$\tau$\\
\hline
1&$1.15^{+0.12}_{-0.17}$&$1.04^{+0.10}_{-0.12}$&$.022^{+0.003}_{-0.004}$&$.53^{+0.25}_{-0.13}$&$<0.85$&$<0.4$\\
2&$1.00^{+0.03}_{-0.02}$&$1.04^{+0.10}_{-0.08}$&$.022^{+0.003}_{-0.002}$&$.69^{+0.08}_{-0.06}$&$0.73^{+0.09}_{-0.07}$&$<0.42$\\
\hline
\end{tabular}
\end{table}

\begin{figure}[t]
\begin{center}
\includegraphics[scale=0.5]{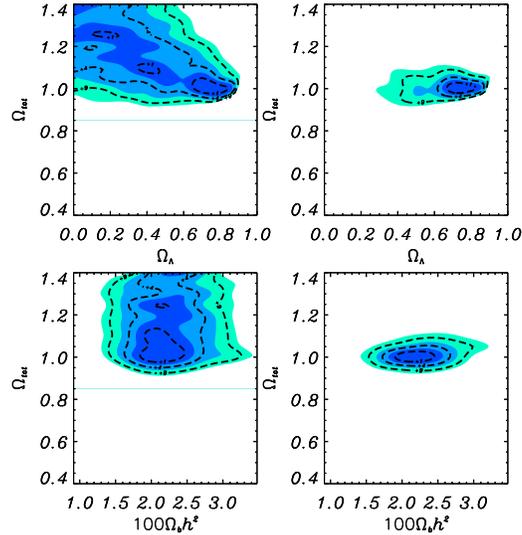}
\caption{Likelihood contours in the $\left(\Omega_{tot},\,\Omega_{\Lambda}\right)$ and $\left(\Omega_{tot},\,\Omega_{b}h^{2}\right)$ planes. Left : constraints using Archeops+CMB data. Right : adding HST prior for $h$.}
\label{contour_cosmo}
\end{center}
\end{figure}

Table \ref{table_param} gives the constrains on cosmological parameters with the data sets Archeops+CMB and Archeops+CMB+HST. Some results are also presented as 2-D contour plots on figure \ref{contour_cosmo}, showing in shades of blue the regions where the likelihood function for a combination of any two parameters drops to 68\%, 95\% and 99\% of its maximum value. Black contours mark the limits to be projected if confidence intervals are sought for any one of the two parameters. 

Note that considering Archeops+CMB data, a strong degeneracy remains, that is well broken
when adding the HST prior on the Hubble constant. The tight constrain on $\Omega_{tot}$, at 3\% level, is compatible with the hypothesis of flatness of the Universe, and the constrain on $n_{s}$ is compatible with the hypothesis that the primordial fluctuations are nearly scale-invariant. These results lend support to the inflationary paradigm. Note also that the constraints on $\Omega_{\Lambda}$ and $\Omega_{b}h^2$ agree well with the results of the Supernovae Ia \cite{sn1a} and Big Bang Nucleosynthesis \cite{bbn}, with a different approach and systematic effects. 
More details can be found in the Archeops paper \cite{papier_cp}.

\subsection{Dust polarization}

The measurement of the CMB polarization spectra, since the pioneering works of DASI \cite{dasipol} and WMAP \cite{wmapol}, has now become the major experimental goal in the field. For high frequency CMB experiments, the major foreground to remove is the emission of the Galactic Interstellar Dust. Its intensity can be inferred from IRAS and COBE/DIRBE extrapolations \cite{schlegel}, but nothing is known on its polarization on scales larger than 10 arcminutes. As these scales are the relevant ones for CMB polarization studies, the determination of the maps of the Galactic Interstellar Dust polarization can be seen as a mandatory preliminary step. 

To achieve this goal, Archeops uses three pairs of polarized bolometers at 353 GHz, a frequency in which the Galaxy dominates. Within each pair, each bolometer observes the same point of the sky, the separation of the incoming light in two orthogonal polarization directions is made by an Ortho Mode Transducer \cite{boifot}\cite{chatt}. 

Maps of the $I$, $Q$, $U$ Stokes parameters over 17\% of the sky and with 13 arcmin resolution have been obtained, with a sensitivity at $1\,\sigma$ per 27 arcminute pixel of 82 $\mu\rm K_{RJ}$ for I, and 105 $\mu\rm K_{RJ}$ for $Q$ and $U$. The results and their implications are presented in \cite{papier_dust}. A dust polarisation of about 5--6 \% is found in the galactic plane on average, and as much as 15--20 \% in a few extended clouds.

\subsection{Prospects}

The recent measurement of the power spectrum of the CMB anisotropies with the WMAP space mission  makes the Archeops measurement of the $C_\ell$ as published obsolete.
Archeops, however, remains complementary to WMAP because it covers a large fraction of the sky in a complementary frequency domain. This opens the possibility of combining WMAP and Archeops for a wide multi--frequency study of astrophysical emissions in the millimeter range.

It is not planned at the moment to fly Archeops again as it is. More sensitive instruments for CMB studies are operational (as WMAP) or in construction (The Planck mission), so that the scientific objectives of Archeops future flights have to be revised.

Two interesting possibilities are being investigated. The first is complementing the sky and frequency coverage available to the scientific community (more sky with Archeops at frequencies complementary to WMAP). The second would be to measure large scale polarisation, and in particular confirm at 150 and 217 GHz the WMAP measurement of a high $T-E$ cross power spectrum.

\section{Conclusion}

The Archeops experiment, flown succesfully three times from 1999 to 2002, has gathered a large amount of high quality data. The data, currently being analysed, has already yielded a good measurement of the CMB anisotropy power spectrum in the angular scale range of the first acoustic peak.

In addition, a measurement of the polarisation of the emission of galactic dust at 353 GHz has been obtained.

Much more is still being expected from Archeops data. In particular, high quality large size maps of sky emission in the four Archeops channels, as well as maps of astrophysical components, are expected to be released in a near future. Combined to other data as maps from the WMAP, from IRAS, and from COBE, these data products will be a precious tool to further our understanding of sky emission in the millimeter range.

\theendnotes

\end{article}
\end{document}